\ifdefined\anonymousbuild
  \documentclass[anonymous, manuscript, screen, nonacm, review, natbib=false]{acmart}
\else
  \documentclass[manuscript, screen, nonacm, natbib=false]{acmart}
\fi
\usepackage[natbibapa]{apacite}
\usepackage{fontspec}
\usepackage{xeCJK}
\IfFontExistsTF{Kaiti SC}{
  \setCJKmainfont{Kaiti SC}
}{
  \IfFontExistsTF{AR PL UKai CN}{
    \setCJKmainfont{AR PL UKai CN}
  }{
    \IfFontExistsTF{Noto Serif CJK SC}{
      \setCJKmainfont{Noto Serif CJK SC}
    }{
      \IfFontExistsTF{SimSun}{
        \setCJKmainfont{SimSun}
      }{
        \setCJKmainfont{FandolKai-Regular.otf}
      }
    }
  }
}
\usepackage{todonotes}
\newcommand{\citepos}[1]{\citeauthor{#1}'s \citeyearpar{#1}}

\usepackage{multicol}
\AtBeginDocument{%
  \renewcommand{\rhead}[1]{}%
  \renewcommand{\lhead}[1]{}%
}

\copyrightyear{2026}
\acmYear{2026}
\acmDOI{}
\acmConference[]{}{}
\acmISBN{}

\begin{document}

\title{Navigating relational agency in AI feedback musicianship: a diffractive dialogue}

\author{Yichen Wang}
\authornote{These authors contributed equally to this research.}
\email{yichen.wang@anu.edu.au}
\orcid{0000-0001-9887-0247}
\affiliation{%
  \institution{The Australian National University}
  \city{Canberra}
  \state{ACT}
  \country{Australia}
}

\author{Charles Patrick Martin}
\authornotemark[1]
\email{charles.martin@anu.edu.au}
\orcid{0000-0001-5683-7529}
\affiliation{%
  \institution{The Australian National University}
  \city{Canberra}
  \state{ACT}
  \country{Australia}
}

\author{David Gunawan}
\affiliation{%
  \institution{Dolby Laboratories}
  \city{Sydney}
\state{NSW}
  \country{Australia}}
\email{dguna@dolby.com}

\author{David Cooper}
\affiliation{%
  \institution{Dolby Laboratories}
  \city{Sydney}
\state{NSW}
  \country{Australia}}
\email{david.cooper@dolby.com}

\begin{abstract}
Musical AI systems behave in non-linear, dynamic ways that exceed the tool-oriented framings inherited from earlier instrument design. 
Existing frames of control and ensemble interaction do not articulate the entangled, relational agency that emerges when performers work with AI feedback instruments over time. 
We examine a duet practice with two AI-enabled feedback instruments, 溯 (Sù) and Agentier, in which feedback operates at the inner level of the AI's recurrent processes and the outer loop of musical interaction between performer, instrument, and co-performer. 
Our work draws on agential realism and a diffractive dialogue methodology to read our independent reflections through one another across five performances, examining how agency is enacted between performers and AI instruments. 
We find that control and agency are relational and shift across performances; musical textures enact \emph{feedback moments} of control with performers; reaching a productive \emph{feedback environment} is itself a shared exploratory act; and performers maintain their instruments through \emph{funnelling} and \emph{injection}, developing the skill of \emph{navigating agency} along a continuous spectrum. 
Feedback musicianship with AI instruments demands navigation of this relational agency, displacing the leading-versus-responding binary that initially framed our reflections. 
This work contributes a materially-grounded vocabulary for AI feedback musicianship and a model of human-AI creative collaboration in which mutual agency is worked for through practice rather than designed in.
\end{abstract}


\maketitle

\section{Introduction}

\begin{figure}
  \includegraphics[width=0.48\textwidth]{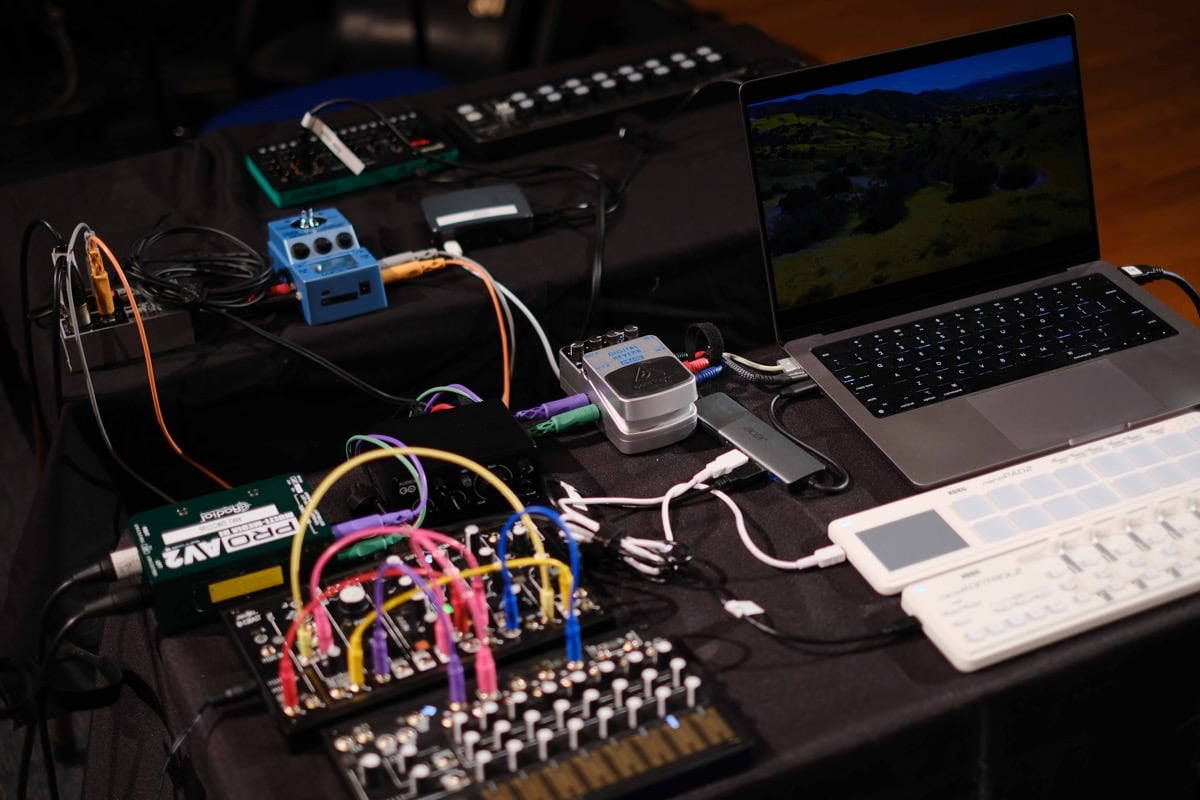}  
  \includegraphics[width=0.48\textwidth]{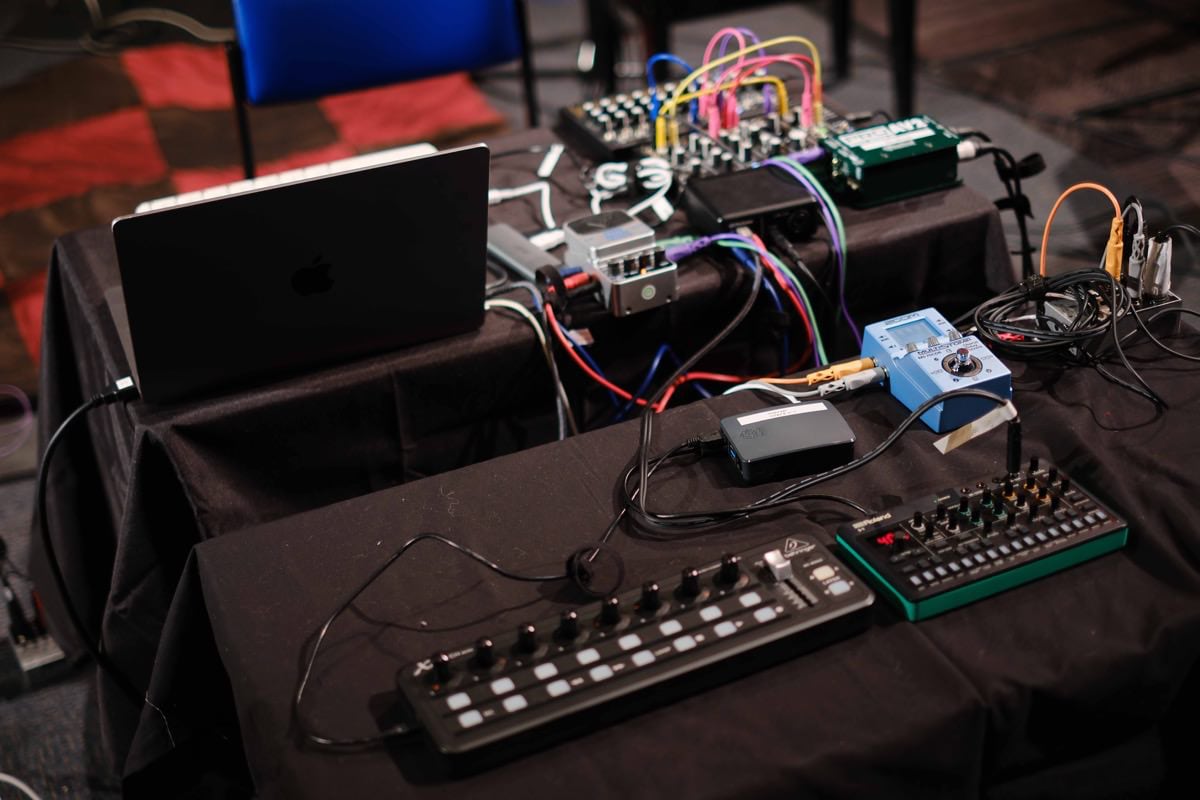}
  \caption{Caption: Our two feedback intelligent musical instruments set for our first performance (P1) in November 2025. All performance data is available online~\anon[\citep{anon}]{\citep{wang-martin-diffractive-data:2026}}: \anon{\url{https://doi.org/10.5281/zenodo.19395011}}\label{fig:smcclablivesetup}\\
  Alt Text: Images of the two instruments in our study on tables in a concert hall under stage lighting. The left image shows a Make Noise synthesiser, Korg nano pad, and fader controllers, and a laptop. The right image shows an xTouch knob controller, Roland S-1 synthesiser, and Raspberry Pi.}
  \Description[Images of the two instruments in our study.]{Images of the two instruments in our study on tables in a concert hall under stage lighting. The left image shows a Make Noise synthesiser, Korg nano pad and fader controllers, and a laptop. The right image shows an xTouch knob controller, Roland S-1 synthesiser, and Raspberry Pi.}
\end{figure}

Two musicians sit at a table recording a performance with feedback AI instruments. One involves an audio feedback loop, filtering and re-creating sound from an audio-generating neural network. The other involves gestural feedback; a neural network generates parameter and note changes feeding back on its output and input from the musician.
In the middle of the noisy performance, a gap appears in the texture, a moment of sparse activity. Both musicians identify this moment as notable; a turning point in the improvisation, but the understanding of our instruments' agency differed. 
One performer reads this as a moment of trust and tight control, the other as a moment where agency is offloaded, to be later regained.

Our experience of performing with feedback AI instruments seems to often lead to this kind of puzzling moment. 
Even though we are participants of the same performance, hearing the same sounds, we read the relationship with our instruments' agency differently, even oppositely, as it appeared in the above example.
In this paper we use the framing of feedback musicianship and the methodology of diffractive reading to help unpack these experiences of ensemble performance with AI musical instruments.
We seek to examine how the instruments take an active and changing role in mediating musical meaning rather than only transmitting their performers' intention.

In our context, the first two authors developed an artistic practice using AI-powered musical instruments between 2024 and 2025. 
What we call intelligent musical instruments are instruments involving AI mechanisms that can act with agency independently of a musician's direct control~\anon[\citep{anon}]{\citep{martin-opening-design-space:2026}}. Our work evolved from a solo performer using a MIDI controller,  augmented reality headset, and AI-powered synthesiser~\anon[\citep{anon}]{\citep{wang-off-the-shelf:2024}} to an ensemble of two humans and one AI instrument~\anon[\citep{anon}]{\citep{wang-ai-see-you-see:2025}}, to the developed duo performance involving two intelligent musical instruments, 溯 (Sù) and Agentier~\anon[\citep{anon}]{\citep{wang2026huisucoconstructingdual}} (shown in Figure~\ref{fig:smcclablivesetup}). 
Our practice overlaps with the concept of feedback musicianship, defined as ``making music that explores the recurrent circulation of signals as a core of musical agency''~\citep{kiefer2025feedbackmusicianship}. The overlap is both technical and practical. Our instruments, like many musical AI systems, involve recurrent circulation of data as the AI generated output relies on contextual information built up from previous input and output, a technical connection with feedback instruments.
Practically, performing with intelligent musical instruments involves shared or entangled agency as described by \citet{feedback:magnusson2022} and \citet{playingfeedback:mudd2023}, as well as a practice which is ``complicated, confusing, and unpredictable''~\citep{playingfeedback:mudd2023} but musically rewarding.

The dual feedback instrument setup, and the series of performances applying this setup makes our specific practice worth examining within the frame of feedback musicianship.
Our instruments both involve AI models and feedback loops, but in different ways. 
溯 (Sù), created by \anon[Author1]{Yichen Wang}, uses a RAVE audio-generating AI model, with feedback taking place with audio signals filtered through the RAVE model's latent space~\anon[\citep{anon}]{\citep{wang2026huisucoconstructingdual}}. 
Agentier (a play on agentic keyboard), created by \anon[Author2]{Charles Martin}, uses a recurrent neural network that generates multiple MIDI signals simultaneously, in feedback with its own previous output or the musician's interactions, so its feedback operates within a space of musical gestures~\anon[\citep{anon}]{\citep{martin-opening-design-space:2026}}.
Contrasting our approaches to performing with these two different but related instruments within the same performances presents an ideal opportunity to find meaning beyond the technical challenges of individual instruments.
The unpredictability of generative AI approaches in our instruments allows us to explore the gap in studying digital feedback instruments identified by \citet{playingfeedback:mudd2023}.

In exploring our practice, we turn to \citepos{meeting:barad2007} post-humanist agential realism, as applied within entanglement HCI~\citep{entanglement:frauenberger2019}. In agential realism, ``things are only coming into existence through discursive practices''~\citep{entanglement:frauenberger2019}, which helps unpack the dynamic, intra-active agency of our AI feedback instruments.
Situated within recent applications of Baradian thinking in DMI research~\citep{morrison2024entangle, loop:gaster2025}, we treat agency as emerging and relational through the co-exploration between performers and instruments.
We ask: \emph{How does agential realism shape our understanding of the relationship between performers and AI instruments in feedback musicianship?}
Given the contrasts between our instruments and that our analysis involves two performers, we apply a diffractive reading approach to sense-making in our practice.
Having recorded five performances, each musician wrote reflections, guided by  questions examining our sense of control, and of the agency of our instruments. 
We then engaged in recorded diffractive dialogues, adopting \citepos{morrison2024entangle} terminology, where we sought points of difference and alignment guided by framing questions.
Finally, we identified five themes within our dialogues that trace our experience of feedback musicianship, and our evolving skills, techniques, and understanding of our instruments.

This paper contributes a vocabulary for the practice of feedback musicianship. We introduce concepts such as \emph{feedback moments}, \emph{feedback environment}, funnelling, injection, and navigating agency that emerged from our diffractive process.
Our work identified relational and co-enacted readings of our instruments' agency where binary framings such as leading versus responding were initially expected.
We offer a practice-based account of development over time where the practice and analytic vocabulary developed across five performances (two concerts and three recordings).
These findings are organised into five themes presented in Section~\ref{sec:results}.
More broadly, our analysis of performance with musical AI systems, where the performers and systems co-adapt~\citep{Mackay:2000aa} in real-time, contributes a model of human-AI creative collaboration where mutual agency is worked-for and developed without short-cuts. 

\section{Related Work}

\subsection{Feedback Aesthetics, Material Agency and Musicianship}

Feedback instrument research has explored varied musical designs and phenomena, from electromagnetic augmented instruments~\citep{lorentzfeedback:schimdt2025,drumfeedback:wolstanholme2025}, to ecologies of  relationships in calligraphy-inspired installations~\citep{collaboration:kuzmin2025}, to the entangled relationships between performers and interfaces~\citep{self:eldridge2024}.
\citepos{feedback:magnusson2022} recent reflections on feedback music practice observed that musicians working with feedback systems exhibit a relational agency shaped by their perceptions.
They questioned the mechanisms through which humans attribute agency to mechanical objects, such as playing with, playing against, or being played by these objects.

In a case study of feedback musicianship reported by \citet{kiefer2025feedbackmusicianship}, Paul Stapleton reflected on practice-based feedback music through the lens of distributed agency among actors in a musical system.
He observed that ``feedback is a necessary quality of any living system\ldots\ once things reach equilibrium, it's death''.
This perspective foregrounds the dynamic autonomy of the instrument and the generative unknowns that emerge in feedback musicianship.

\citet[p. 7]{playingfeedback:mudd2023} argued that feedback leverages non-linear, dynamic interaction and encourages ``letting go of intention''.
This connects with entanglement HCI~\citep{entanglement:frauenberger2019}, which emphasises the co-produced relationship between human and machine agency rather than considering agency separately on either side.
\citet{playingfeedback:mudd2023} notes that feedback is less fully explored in a digital musical instrument context, a gap that we explore in this research.
From a material engagement perspective, musical instruments are understood as active mediators in the creative process, capable of shaping ideas~\citep{mudd2019material}.
Musical phenomena emerge through the interaction between performer and instrument. 
This contrasts with communication-oriented perspectives framing musical expression as the transmission of a musician's intention through an instrument.

The incorporation of AI into feedback instruments  introduces further possibilities for material agency within feedback practice.
\citet{kiefer2025feedbackmusicianship} noted: ``AI systems such as autoregressive models naturally encapsulate feedback mechanisms, creating nested feedback loops across hybrid materials with potential for complex behaviours and rich performance experiences''. 
Examples include the folk-inspired Proto-Langspil monochord~\citep{armitage2022protolangspil}, the Feedback Feedward shared instrument~\citep{feedbackforward2021}, the Xiasri AI-enabled string instrument~\citep{xiarsi2022}, and Sophtar~\citep{sophtar:visi2024}, an embedded AI string instrument.
\citepos{sophtar:visi2024} Sophtar drew on artistic practices in feedback guitar and interactive machine learning, with Visi incorporating familiar musical gestures to navigate timbre in an audio-generating RAVE~\citep{rave:caillon2021} model's latent space.

Despite the alignment between AI and feedback instruments, the material agency afforded by AI models' own recurrent processes has received limited attention. 
Sophtar demonstrates how RAVE can be incorporated into feedback instrument design~\citep{sophtar:visi2024}, yet the theoretical implications of nested feedback loops, where AI inference processes constitute an inner feedback layer entangled with the outer performative loop, remain underexplored. 
In particular, how these hybrid feedback structures co-produce musical agency, rather than simply serving as sound-generation tools, is an open question. 
Our work takes up that question directly, treating two AI-enabled feedback instruments as an apparatus through which material, computational, and performative agencies are examined.

\subsection{Intelligent Musical Instrument Design and Musical Practice}

Intelligent musical instruments, or musical instruments that apply machine learning or artificial intelligence techniques to operate autonomously, in dialogue with, or to assist performers have been explored for decades.
Early work on AI composition~\citep{roads-music-AI-1985, Cope:2001aa} led to use in performance~\citep{Lewis:2000fu, Pachet:2003wd, Biles:2007aa}, and in interpreting sensor data~\citep{fiebrink_meta-instrument_2009, fiebrink_years_ml}.
Intelligent musical instrument designs often involved a co-adaptive~\citep{Mackay:2000aa} feedback loop between musician and instrument; musicians creating new instruments adapt their performance practice while updating the instrument design and the learning model.

More recent AI techniques enabled intelligent musical instruments to learn and generate more complex data. 
This data could relate to embodied musical gestures~\citep{Martin2019}, or physical controller movements~\citep{martin_empi}, as is the focus of the system applied in this research.
Audio generating systems such as the RAVE variational autoencoder~\citep{rave:caillon2021}, have been applied within intelligent musical instruments, such as the Living Looper~\citep{livinglooper:shepardson2025}.
A similar audiovisual autoencoder was used in Walshe and Akten's ULTRACHUNK~\citep{dyer-walsh-intelligent-machines:2023}.
In reaction to big-data industrial AI, small-data approaches have emerged as an aesthetic and practical  choice~\citep{Vigliensoni:2022}.
Practice-based accounts of intelligent musical instrument performances are now emerging that engage with the agency of these systems. 
\citepos{stefannsdottir2025altered} account of performing with the Living Looper and violin tracks the process of training and engaging with the RAVE model while framing the violin and AI sounds separately as ``two forms of human-technology relations\ldots\ unfolding simultaneously''.
\anon[The second author's previous work~\citep{anon}]{Martin's previous work using IMPSY~\citep{martin-opening-design-space:2026}} focussed on prototyping and iterating on intelligent musical instrument designs but acknowledged the limitation of a single performer's perspective.
Practice-based accounts have treated these two kinds of AI separately: audio AI as enacted in the performance space and gestural AI as operating in the embodied space of the performer.
Our duet practice analyses both simultaneously.

\subsection{Diffractive Approaches in DMI and Creative HCI Research}

In the post-human turn to entanglement in HCI~\citep{entanglement:frauenberger2019} and DMI research~\citep{morrison2024entangle} diffractive approaches have been adopted to explore the intricate relationship between musicians and their instruments.
Drawing on feminist theorist and physicist Karen Barad, diffraction is understood as a method for ``reading insights through one another in attending to and responding to the details and specificities of relations of difference and how they matter''~\cite[p. 11]{meeting:barad2007}.
Diffraction allows us to produce new understandings of technologies by examining multiple perspectives in relation to one another, rather than in isolation.

\citet{morrison2024entangle} staged a diffractive dialogue to reflect on DMI design, interleaving their perspective with knowledge from music, media, and cultural studies.
\citet{loop:gaster2025} adopted a diffracted reading to explore how both old and new technologies are used in musical practices.
\citet{materialpolitics:allen2025} frame a hair-braiding-inspired DMI as a Baradian apparatus to surface the politics of design knowledge in NIME.

Diffractive approaches have also been applied to  creative human-AI interaction, as a lens for what can be an ``ill-defined, iterative and exploratory process''~\citep{creativeAIdiffraction:bird2025}.
\citet{ritual:rajcic2023} adopted diffractive analysis to examine how an AI-enabled ambient lamp ``becomes with'' participants, revealing how AI mediates memory and meaning through ongoing, everyday interaction.
An agonistic dialogue, similar to \citepos{morrison2024entangle} diffractive dialogue, was applied by \citet{Kaila2024Agonistic} to interrogate AI music systems.

These examples have shown diffraction used to cope with the multiple perspectives involved in DMI design, and the mutual shaping and ongoing negotiation of human-AI interaction.
Situating diffractive analysis within this work, we seek to diffract not only our individual perspectives, as has been done in previous work, but the material agency of our two distinct AI-enabled feedback instruments.
We conceptualise our performance practice as a material-discursive process, examined by diffraction, generating meaning about the design of intelligent musical instruments.

\section{Apparatus and Performances}

\begin{figure}
\includegraphics[width=0.95\textwidth]{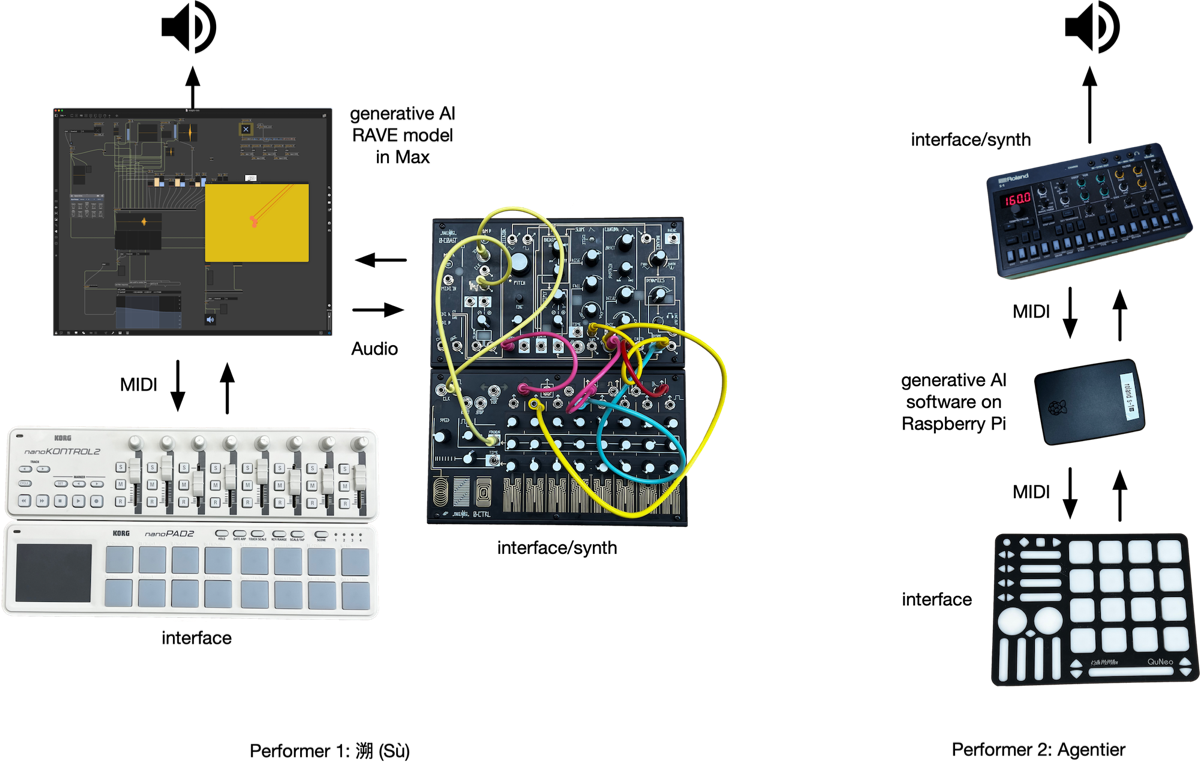}
\caption{
  Caption: Stage setup diagram showing signal flow for the two instruments.\\
  Alt Text: A signal diagram for devices in our two instruments. Sù (left): a Max patch is connected with audio in and out to a Make Noise 0-COAST and 0-CTRL synthesiser setup, and with MIDI to Korg nanoKONTROL and nanoPAD controllers. Audio from the Max patch is sent output to the stage. Agentier (right) has a Raspberry Pi connected by MIDI to a Keith McMillen QuNeo pad controller and a Roland S-1 synthesiser which sends sound output to the stage.}\label{fig:signal-diagram}
\Description[A signal diagram for devices in our two instruments]{Sù (left): a Max patch is connected with audio in and out to a Make Noise 0-COAST and 0-CTRL synthesiser setup, and with MIDI to Korg nanoKONTROL and nanoPAD controllers. Audio from the Max patch is sent output to the stage. Agentier (right) has a Raspberry Pi connected by MIDI to a Keith McMillen QuNeo pad controller and a Roland S-1 synthesiser which sends sound output to the stage.}
\end{figure}

This work involves a multilayered collaborative apparatus centred around two intelligent musical instruments, 溯 (Sù) and Agentier. 
In the Baradian sense, this apparatus co-produces musical phenomena with performers rather than merely transmitting intention~\citep{morrison2024entangle}.
Both instruments involve feedback loops at the level of AI-inference, between musician and instrument, and between the two musicians. In 溯, AI feedback operates within the audio domain, while in Agentier feedback is modelled in the gestural (MIDI data) domain. The stage setup for our instruments is shown in Figure~\ref{fig:signal-diagram} In this section we describe these instruments to motivate our later performance research analysis.

\subsection{溯: Enacted Feedback Control in the Audio Space.}

\begin{figure}
\includegraphics[width=0.7\textwidth]{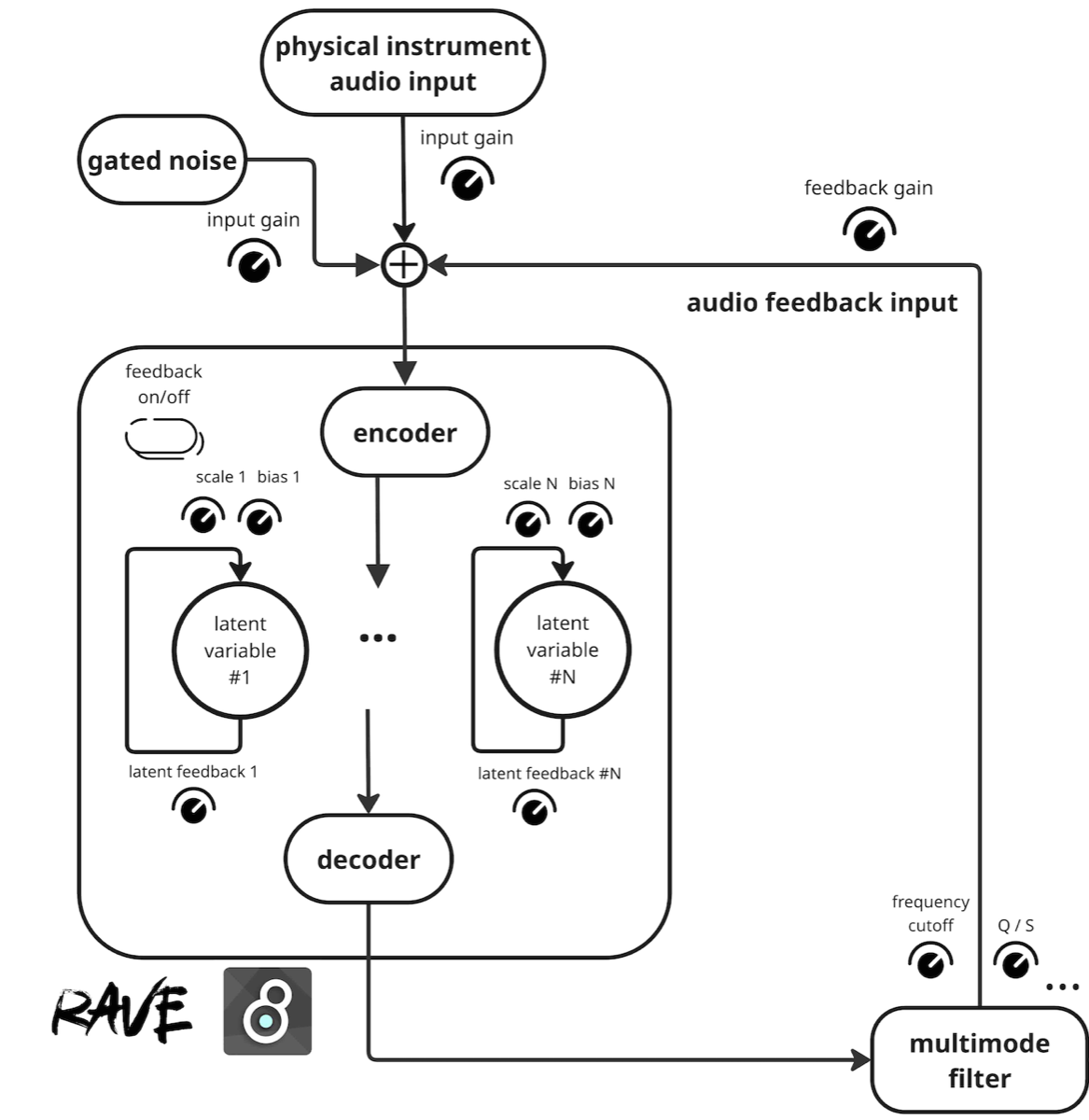} 
\caption{Caption: The system diagram for 溯 (Sù) showing the outer audio feedback loop, and inner loop of adjustable feedback of latent variables between the RAVE encoder and decoder.\\
Alt Text: A system diagram for Sù showing inputs and outputs to the feedback loops. The outer audio loop takes audio from the RAVE decoder, goes through a filter and gain controller, mixes with input from the physical instrument and gated noise and returns to the RAVE encoder. Inside the RAVE model, each latent variable is biased by controllable feedback from its previous value.}
\label{fig:su-system-diagram}
\Description[A system diagram for Sù showing inputs and outputs to the feedback loops.]{The outer audio loop takes audio from the RAVE decoder, goes through a filter and gain controller, mixes with input from the physical instrument and gated noise and returns to the RAVE encoder. Inside the RAVE model, each latent variable is biased by controllable feedback from its previous value.}
\end{figure}

溯 (Sù, to trace back; to go upstream) is a novel latent feedback instrument built on a pretrained RAVE~\citep{rave:caillon2021} vocal model \texttt{Isis}~\citep{ravemodels:acids} implemented in Max/MSP.
As shown in Figure~\ref{fig:su-system-diagram}, the instrument integrates three interacting mechanisms: audio feedback, latent feedback, and direct latent manipulation.
The audio feedback mechanism is introduced as a controllable gain that feeds the model's audio signal back into its input.
This signal can be combined with an external hardware synthesiser and noise source, allowing richer sonic materials to circulate within the feedback structure.
A second feedback loop is introduced within the RAVE model's inference process.
For each latent dimension, the encoder's latent parameters at the current audio window are combined with the corresponding parameters from the previous window.
This latent feedback biases the current latent distribution toward the previous state, introducing temporal dependency and reducing abrupt changes between successive windows.
In parallel, individual latent dimensions can be directly manipulated via MIDI control prior to decoding, allowing performers to shape the timbral characteristics of the generated sound.
These manipulated dimensions remain subject to latent feedback, enabling timbral intervention while preserving temporal continuity.

The instrument responds to limitations in neural synthesis systems that rely on learned latent space representations (e.g., timbre features), which offer limited control over higher-dimensional sonic features for creative expression~\citep{latentdace:nabi2024}, despite their growing use in performance practice~\cite[e.g.,][]{stacco:privato2024,latentroom:shaheed2024}.
Rather than relying on conditioning or assumptions about learned representations~\citep{latentdace:nabi2024}, 溯 borrows the recursivity of the feedback technique to stabilise navigation in the latent space, supporting more controllable timbral behaviour in performance.

\subsection{Agentier (Agentic Clavier): Embodied Feedback in the Gestural Space.}

Agentier uses a novel musical AI platform to mediate and generate gestural information (MIDI messages) between two performance interfaces.
An embodied musical gesture AI model using an auto-regressive mixture density recurrent neural network (MDRNN)~\citep{Martin2019} was trained on data from eight continuous controllers, recorded by the second author.
The model can generate and continue streams of musical control data including the time at which updates should be performed.
This model works in a call-and-response loop with the performer with a fast switchover time of 0.1 seconds. In practice, this means that the model can fill in between the performer's gestures and can be guided and mediated by continual interactions.
Two interfaces, a synthesiser and touch-pad controller are connected to the AI model, with input received from and output sent to both simultaneously. The eight control outputs are mapped to notes and seven timbral parameters on the synth and eight LED touch sliders on the touch-pad. Inputs from the touch-pad's sliders and pads, and the synth's knobs and keys map to the 8 inputs of the model.
The two interfaces form an assemblage~\citep{musical_instruments_as_assemblage_theberge} with attributes of each interface contributing to the new intelligent instrument.

\subsection{Performances}
\label{sec:performances}

\begin{figure}
\includegraphics[width=0.7\textwidth]{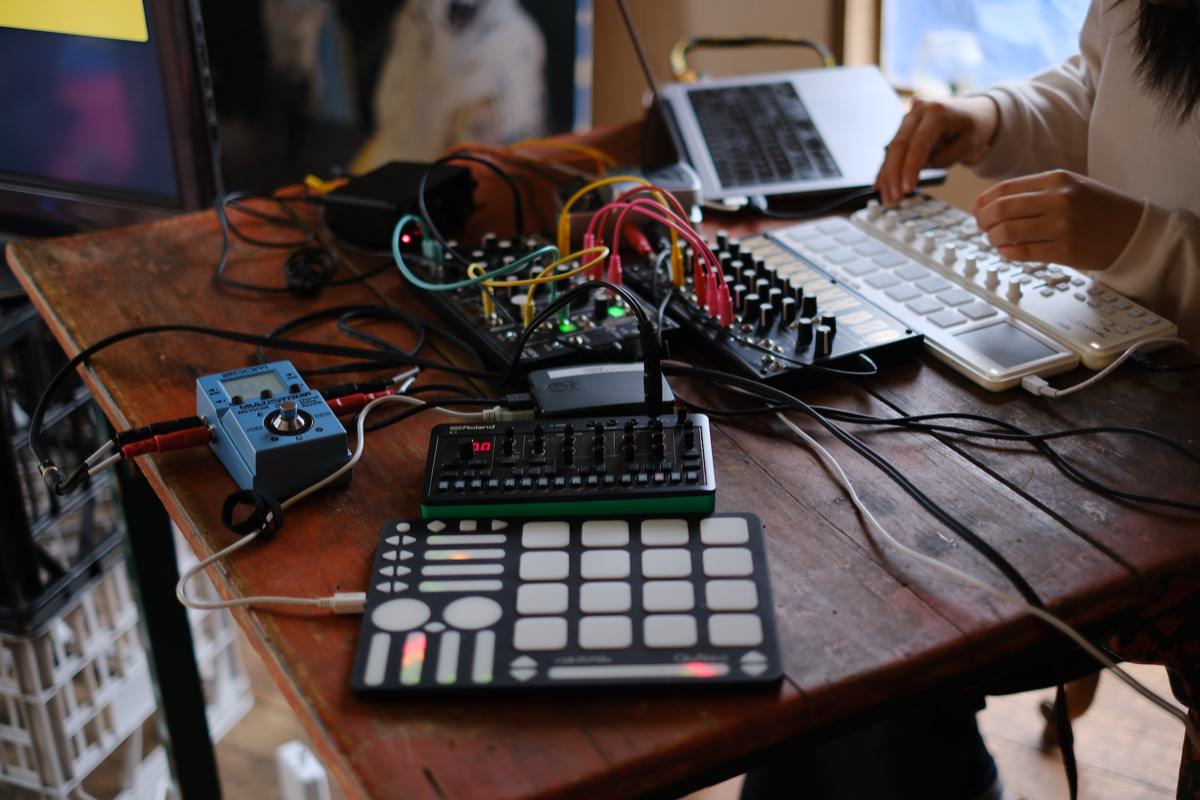} %
\caption{Caption: \anon{Deathless Wonder} performance (P2)\\
Alt Text: Our setup at the DIY backyard performance. Sù and Agentier set up on a rustic table with natural light. A woman performs with Sù. Milk crates can be seen in the background.}
\label{fig:deathlesswonder}
\Description[Our setup at the DIY backyard performance.]{Sù and Agentier set up on a rustic table with natural light. A woman performs with Sù. Milk crates can be seen in the background.}
\end{figure}

\begin{figure}
\includegraphics[width=0.7\textwidth]{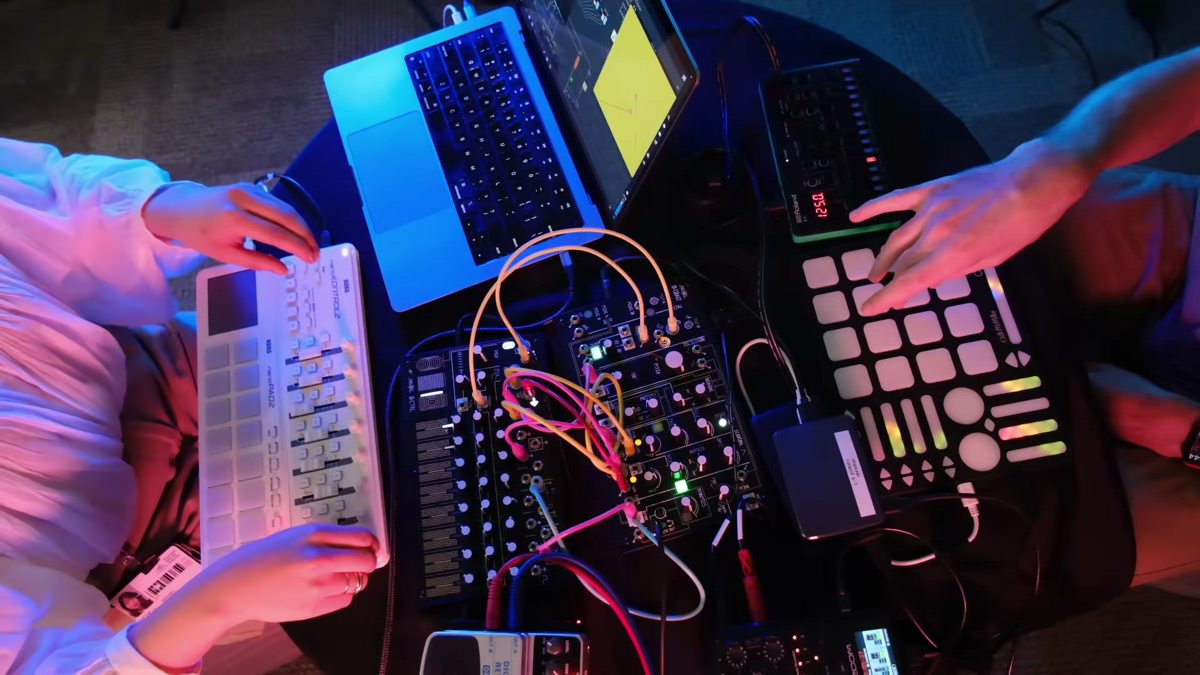} %
\caption{Caption: Recording setup for P3--5\\
Alt Text: Sù and Agentier set up on a round table. The camera is above the instruments and the room is dark except for coloured lights. A woman performs with Sù and a man performs with Agentier.}
\label{fig:recording}
\Description[Sù and Agentier set up on a round table.]{The camera is above the instruments and the room is dark except for coloured lights. A woman performs with Sù and a man performs with Agentier.}
\end{figure}

\begin{table*}[]
\caption{Configuration of the performances, recordings, and instrument configuration considered for our feedback practice. Our analysis considers five separate recordings including two live performances and three recorded takes. This corpus included 87 minutes of recorded performances. All performance recordings are available online~\anon[\citep{anon}]{\citep{wang-martin-diffractive-data:2026}}.}
\label{tab:perf-corpus}
\begin{tabular}{@{}llllll@{}}
\toprule
Code & Date     & Type      & 溯 (Sù) setup & Agentier setup & Duration (min) \\ \midrule
P1 & 2025/11/03 & Concert (concert hall) & 溯 + percussion model & S-1 + xTouch + \anon[MDRNN]{IMPSY} & 23 \\
P2 & 2025/11/22 & Concert (DIY backyard) & 溯 + voice model & S-1 + QuNeo + \anon[MDRNN]{IMPSY} & 29 \\
P3 & 2025/12/19 & Recording & 溯 + voice model & S-1 + QuNeo + \anon[MDRNN]{IMPSY} & 11 \\
P4 & 2025/12/19 & Recording & 溯 + voice model & S-1 + QuNeo + \anon[MDRNN]{IMPSY} & 12 \\
P5 & 2025/12/19 & Recording & 溯 + voice model & S-1 + QuNeo + \anon[MDRNN]{IMPSY} & 12 \\
\bottomrule
\end{tabular}
\end{table*}

The artistic aspects of this research involved five performances in November and December 2025 (listed in Table~\ref{tab:perf-corpus}). The first two performances were live recordings at concerts. The second three were three video recordings made in one day.
Performance 1, in a research lab concert (\anon{\emph{smcclablive}}, Figure~\ref{fig:smcclablivesetup}), took place at the \anon{ANU Big Band Room} concert hall with video and audio recording. Performance 2 was part of a DIY music series (\anon{\emph{Deathless Wonder}}, Figure~\ref{fig:deathlesswonder}) performed in an \anon{inner-Canberra} suburban backyard with a small PA system and was audio recorded only. The recordings (Figure~\ref{fig:recording}) were created in a university office and were intended as artistic development and documentation of our performance practice.

The context, instrumental setup, and level of documentation changed in between the performances. The first involved an earlier version of Agentier where a knob controller was used, this was replaced with a touch-pad controller for the later performances. The Sù setup also varied with the first performance involving a RAVE percussion model which was changed to a voice model for later sessions.
P1 was live-streamed with video recorded from the front of the stage overlaid on a screen recording of the Max patch from Sù. The second performance was not video recorded. The last three performances had a top down camera recording the movements of both musicians with another camera angle angled across the instrument setup.

\section{Methodology}

To explore what feedback reveals about the co-constructed relationship between musicians and intelligent musical instruments, we adopted a diffractive dialogue approach, adapted from \citet{morrison2024entangle}.
Our analysis involved the following phases: individual reflections on performance, a diffractive dialogue which we read our reflection through each other's insights, and collaborative analysis.
These phases allow us to examine how our experiences interfere, overlap, and diverge in understanding the emerging and relational agency with intelligent musical instruments.
This process involved the first two authors who were also the performers in Section~\ref{sec:performances}.

\subsection{Phase 1: Individual Written Reflection}

In this phase, we individually reviewed our performance recordings and wrote reflections following four prompts.
The prompts were adapted from interview questions in \citet{playingfeedback:mudd2023}, which examined the entangled agency in feedback interaction when performing with no-input mixer. 
We extend these questions to address interactions with our feedback instruments emphasising live performance, the perception of musical control and intention, and key moments that could reveal dynamic interaction.
The prompts seek to expand our understanding of the relationship between performers and AI instruments in feedback musicianship.
They elicit reflection across three levels of feedback mechanisms: within the instruments, between the performers and their instruments, and holistically between the performers.
\begin{enumerate}
  \item How did the feedback mechanism(s) in the performance shape your musical decisions during the performance?
  \item Describe moments where the sense of control felt distributed, dissolved, or unexpected. How did that shift your engagement?
  \item What were your thoughts and intentions at key moments, and how did the performance or instrument behaviour align or diverge from them?
  \item When did you feel the instrument(s) were ``leading'' rather than responding?
\end{enumerate}

\subsection{Phase 2: Diffractive Dialogue}

In the second phase, we engaged in a recorded discussion. We read each other's reflections, then discussed our different perspectives guided by the following prompts:
\begin{enumerate}
  \item Where do our experiences diverge, and what does that difference reveal about each instrument's agency?
  \item Where do our experiences interfere or overlap, and what does that suggest about the shared feedback environment?
  \item Were there moments where our intentions became indistinguishable from the instrument's behaviour, did experiences differ in these moments?
  \item How did our two feedback systems entangle with each other during the performance?
\end{enumerate}

\subsection{Phase 3: Collaborative Analysis}

\begin{table*}[]
  \caption{The reflections and dialogues completed as part of this research. The results were derived from diffractive reading and discussion of the reflections. This corpus comprised 23206 words inclusive of more than 2 hours of discussion and is available online~\anon[\citep{anon}]{\citep{wang-martin-diffractive-data:2026}}.}
  \label{tab:reflection-corpus}
\begin{tabular}{@{}llll@{}}
Code & Reflection          & Duration (time) & Length (words) \\ \midrule
\anon[RA1]{RY} & \anon[Author1]{Yichen} Reflections  &               & 2906           \\
\anon[RA2]{RC} & \anon[Author2]{Charles} Reflections &               & 3577           \\
D1 & Discussion 1        & 34:38         & 4595           \\
D2 & Discussion 2        & 37:06         & 4792           \\
D3 & Discussion 3        & 22:02         & 2705           \\
D4 & Discussion 4        & 19:51         & 2404           \\
D5 & Discussion 5        & 18:59         & 2227           \\
\end{tabular}
\end{table*}

Following the diffractive dialogue we transcribed the audio recording\footnote{Dialogues were transcribed automatically using local transcription and diarisation models: \anon{\url{https://github.com/smcclab/labscriber}}} and collaboratively developed findings drawing from all reflections and dialogues shown in Table~\ref{tab:reflection-corpus}.
We analysed the dialogue transcript to identify patterns and tensions across our experiences.
We identified themes focussed on interferences that would not have been produced alone through our reflections.

\section{Results}
\label{sec:results}

In this section we report five themes identified across our whole corpus, including the diffractive dialogues and individual reflections.

\subsection{Control is a Relational Construct}
\label{theme:control-relational}

Throughout our discussions, we discovered that \emph{control} often meant something different to each performer, and that this meaning changed across the performances.
In our earlier discussions, we identified two different interpretations of being in control within our feedback systems: control identified with a feeling of stability, and control identified with divergence:
\begin{quote}
``you reflected on feeling in control when the voice was changing\ldots you identified control with divergence in the musical output and not with stability, which is kind of the opposite of how I was feeling about control. I was thinking of control as stable moments and divergences.'' (\anon[Author2]{Charles}, D2, L22--31)

``for me, you are trying to put down your system. And for me, it was trying to support my system.'' (\anon[Author1]{Yichen}, D2, L34)
\end{quote}
In practice, the stability of our feedback instruments was different. Sù needed to be cajoled into exploration, while Agentier tended towards excessive divergence. \anon[Author1]{Yichen} felt more in control when Sù was varying its output, while \anon[Author2]{Charles} felt in control when Agentier was not. 
We understood control as trusting that our feedback loops won't explode, versus trusting them not to disappear and these understandings were in relation to the properties of our instruments. 
We had earlier defined these performance strategies as ``maintenance'' (\anon[Author2]{Charles}, D1, L213) which could be ``more supportive'' or ``more restrictive''.

This argument is complicated by the fact that our positions reversed in the final discussion.
In P5, \anon[Author1]{Yichen} felt a sense of control when ``it was in a stable state'' (\anon[Author1]{Yichen}, D5, L76), noting that \anon[Author2]{Charles}' sense of control was ``intentionally trying to make it [move] forward'' (\anon[Author1]{Yichen}, D5, L75):
\begin{quote}
``in the earlier dialogues, it seemed to be that \ldots you wanted more variety and I wanted more stability. But now\ldots I wanted more variety and you wanted more stability. So, through this performance process, you were saying control equals divergence, not stability, whereas I was thinking control equals stability, not divergence. And now we've established the opposite role.'' (\anon[Author2]{Charles}, D5, L89--93)
\end{quote}
As we observed in D5 (L97), our expectations and intentions for the systems changed over time, along with our feeling of control, even though the instruments themselves didn't change their behaviour.

In these performances, our feeling of being in control was relational to the instrument, the co-performer, and our experience and intentions within each performance. 
While both Sù and Agentier involved feedback loops, their different properties led to different feelings of control which were not fixed, but rather, varied with our performative experience.

\subsection{Musical textures enact kinds of control with performers in feedback moments}
\label{theme:textures-enact}

In reflection and discussion we tended to associate different musical textures with different feelings of control. In particular, dense or thick textures tended to lead to a feeling of dissolved or diffuse control where feedback processes were allowed to evolve more freely.
\begin{quote}
``\ldots in a part of the performance where we felt that energy started to expand and the texture got really thick, then I could feel more trusting of my instrument and feel like I could start to let go of the control a little bit while still knowing that it was producing a good outcome'' (\anon[Author2]{Charles}, D2, L124)
\end{quote}
For \anon[Author2]{Charles}, moments with thicker or denser musical texture enabled more trust in the feedback process, and thus a more diffuse sense of control. It became clear that developing a denser texture through, for example, reverb and delay, was a way to enter a \emph{feedback moment}, a state where continuing feedback processes allowed the musician's role to recede:
\begin{quote}
  ``in your reflection, [I] highlighted\ldots a more dissolved feeling of control, which is when having denser textures\ldots you felt less pressure to make constant decisions''. ``I consider actually that would be\ldots you're buffering through reverb and delay, maybe the way that you used to manage this instrument helps it behave in a good positive way. And I actually wonder, would that actually be a good feedback state or considered to be a feedback state between the performer and the instrument?'' (\anon[Author1]{Yichen}, D3, L41--43)

  ``So it's using audio feedback\ldots to manage or to help with the management of the other feedback processes.'' (\anon[Author2]{Charles}, D3, L44--47)
\end{quote}
The reflection that, in these states, multiple kinds of feedback are used to manage each other supports that idea of diffusing control between different processes.

Sparse textures, too, led to notable, but different, feelings of control. A particular moment 8 minutes into the second recording (P4) was read as applying tight constraint by \anon[Author2]{Charles} but as offloading agency by \anon[Author1]{Yichen}. In \anon[RA2]{RC}, \anon[Author2]{Charles} said this moment indicated ``tight control over both instruments'' which ``reveals some trust in our systems, that when we turn them up, the sound is good and useful and within expectations'' (\anon[RA2]{RC}, L175). This led to discussion in D4:
\begin{quote}
``I actually wonder what did you mean by very tight control? Do you mean lots of activities, just engaging with our own instruments?'' (\anon[Author1]{Yichen}, D4, L30--33)

``I meant\ldots the musical output at that point was a very sparse texture, so there was not a lot of sound. There was space between the sounds, and the sounds themselves were controlled in terms of their timbral parameters\ldots it was unusual in that way because often the sounds felt a little bit like a bit more noisy or out of control in general throughout performances\ldots" (\anon[Author2]{Charles}, D4, L34--38)
\end{quote}
In contrast, \anon[Author1]{Yichen} noted this section as a moment where she needed to ``regain control\ldots, mediate / probe the systems behaviour\ldots to reestablish influence'' (\anon[RA1,]{RY,} L122). To \anon[Author1]{Yichen}, this section felt like a loss of control where agency had to be offloaded to the instruments:
\begin{quote}
``Maybe offload. I think this can't be reproduced or I would consider that as offloading some of our agency. Maybe that's where something was working in the feedback environment.'' (\anon[Author1]{Yichen}, D4, L50--52)
\end{quote}
This moment, where trust in the feedback instruments connected to potential offloading of agency also aligned with moments where the performers had, in earlier performances, felt out of control. In D1, \anon[Author2]{Charles} noted that ``out of control or borderline out of control moments'' were ``good because that indicated you were right on the edge of special good behaviour'' (\anon[Author2]{Charles} D1, L226--228).

Through our discussion we identified that notable musical textures, both dense and  sparse, afforded specific types of control. 
In these sections, the feedback mechanisms shaped the performers' sense of control especially clearly so we called these \emph{feedback moments}. 
The textures here emerge from entangled behaviours of performers and feedback instruments. 
In that sense, the musical texture and performers have co-agency to enact kinds of control within feedback moments.

\subsection{Entering the feedback environment is shared exploration}
\label{theme:shared-exploration}

While our improvised performance practice was, by definition, a shared exploration, we identified through discussion that entering the feedback environment was also an exploratory task.
Both instruments' feedback mechanisms were stateful and needed to be built up from scratch during each performance. 
A consequence of this was that some aspects of feedback performance could not be accessed early in the improvisations, and that the process of exploration was necessary to achieve those moments.

The beginnings of performances were described as unstable.
In these phases, instruments felt out-of-control and didn't readily support initial exploration:
\begin{quote}
``I thought the first third of the performance was framed around unexpected intrusions from the two instruments which seemed a bit out of control\ldots I wonder whether this is because of being in a starting state (both instruments and performers) or whether the design of the instruments didn't support initial exploration well.'' (\anon[RA2,]{RC,} L92--98)
\end{quote}
In these situations, the performers explored strategies to activate their instrument's behaviours and bring the instruments to more useful states.
For example, \anon[Author1]{Yichen} reflected how she attempted to prime the instrument's state by dialling up audio input from the 0-Coast synthesiser and gradually handing over to the AI model:
\begin{quote}
``At the beginning, I started by outputting sound from the 0-Coast, then gradually dialled it down while increasing the mediated/transformed output from the AI model\ldots My musical decisions were therefore shaped by exploring how different inputs would activate the model's behaviour, although I was less focused on triggering specific sounds and more interested in how the generated sound evolved over time.'' (\anon[RA1,]{RY,} L15--19)
\end{quote}
Rather than entry into the feedback environment being engineered or predefined, performers and instruments co-produced the conditions of interaction as part of an artistic act.

In discussion, we acknowledged a productive state, the feedback environment, within performances where we were able to more freely understand and influence the feedback process.
The term feedback environment did not appear in our individual reflections, it emerged through discussion as a name for the productive state we both recognised but had previously described differently.
\begin{quote}
``something that we were both thinking about was the discovery or exploration of how to influence that feedback process. So we were both searching for a way to search and guide the process, particularly at the start of the performance, to get it into a state that we felt was productive.'' (\anon[Author2]{Charles}, D1, L64--65)
\end{quote}
Further, given that both instruments had independent and differing feedback mechanisms, we questioned whether it is ``possible or useful to expect that you'd start in a good state right at the beginning'' (\anon[Author2]{Charles} D2, L142--143).
\begin{quote}
``Is it better to accept that the state is random at the start and that you need to spend time getting into a good state or should we kind of start from a preset?'' (\anon[Author2]{Charles}, D2, L146)

``I support the first one\ldots the natural part, like how it just let it start, like evolve. I think particularly we're dealing with an, we assumed, I mean all instruments have their own agency, right? So the way we work with it and trying to get the end seems to be like a whole feedback process.'' (\anon[Author1]{Yichen}, D2, L147--148)
\end{quote}
This process of finding the feedback environment was seen not just as necessary, but also rewarding and useful:
\begin{quote}
``We want to have that feeling that there's that we're building something from scratch and there's the risk that somehow it could all go wrong\ldots feels like a worthwhile artistic act.'' (\anon[Author2]{Charles}, D2, L229--232)
\end{quote}
Given the artistic rewards, we used this process as a marker within our performances:
\begin{quote}
``I would consider that maybe we kind of got into the feedback environment better this time.'' (\anon[Author1]{Yichen}, D3, L26)

``So a sooner and better integration in the feedback environment.'' (\anon[Author2]{Charles}, D3, L27)
\end{quote}

Improvised ensemble performance is often defined by shared exploration, but in our practice, the stateful nature of feedback instruments revealed a different kind of exploration, that of understanding and influencing the feedback mechanisms. We came to see this process as both necessary and desirable. Locating ourselves within the feedback environment emerged as a defining moment in the later performances (P3, P4, P5).

\subsection{Instrumental agency can be maintained through funnelling and injection}
\label{theme:agency-maintenance}

Our reflections and discussions surfaced strategies that we used to guide and maintain our instruments' actions. In other words, to support their agency as feedback instruments participating in our performances. 
\anon[Author1]{Yichen} \emph{injected} noise to wake up the feedback loop driving the RAVE model, \anon[Author2]{Charles} \emph{funnelled} his instrument (that might otherwise escape) by providing constant input, filling up the RNN model's history.
The diffractive dialogue exposed that while our strategies were different and appeared quite opposite in their approach, they served the same function of maintaining our instruments' feedback context, guiding agency in our instruments through support or constraint.

These strategies were initially named in reflections where we, individually, identified and named these ideas. \anon[Author1]{Yichen} framed \emph{injection} around recovering from an out of control moment:
\begin{quote}
``At 04:40---I was unsure whether the model was responding to the input sound from the 0-coast, I `injected' a noise tap to try to `reset' the system, to disrupt the AI model processing the incoming audio. Based on my prior experience, I knew the model would still produce a vocal-like sound with this noise tap, which I could then further shape through adjustments to the latent dimensions." (\anon[RA1]{RY}, L91)
\end{quote}
While \anon[Author2]{Charles} framed \emph{funnelling} around managing moments where Agentier was appearing to lead rather than respond:
\begin{quote}
``I guess I felt my instrument leading whenever the volume was up as it always drags me into different sounds. Rather than responding to me I feel either in direct control or managing the evolving AI process. I feel more like a kind of railing or funnel for the AI system which is constantly trying to go in different directions.'' (\anon[RA2]{RC}, L52)
\end{quote}

Through diffractive reading in D1 the difference between these behaviours became clear and we identified these techniques as maintaining our instruments' behaviour.
\begin{quote}
``Yes, because my instrument is passive, like\ldots You only can let it trigger a sound when it gave some input, right? \ldots and then by giving that input I kind of gave it like found out some, search for some behaviors, right? And then trying to maintain that.'' (\anon[Author1]{Yichen}, D1, L206)

``Okay, so because you're injecting energy into the feedback loop, and then the role is maintenance.'' (\anon[Author2]{Charles}, D1, L209--211)

``Yeah, yeah. So maintenance is first, right? Trying to figure out its behaviour and then see how maybe it can evolve by itself, yeah.'' (\anon[Author1]{Yichen}, D1, L212)

``So I think we both engage in maintenance of a kind, but it's more, yours maybe is more supportive, whereas mine's more restrictive.'' (\anon[Author2]{Charles}, D1, L213)
\end{quote}

In the first discussion (D1), we focussed on the differences between the maintenance strategies, specifically that \anon[Author1]{Yichen}'s seemed to be supportive, while \anon[Author2's]{Charles'} was restrictive. 
By a later discussion (D4), we had clarified the alignment between our approaches and saw more similarities.
\begin{quote}
``\anon[Author2]{Charles} mentioned that in this performance the feedback mechanism to him is trying to overwhelm the RNN loop in the \anon[MDRNN]{IMPSY} system and keep a specific sound\ldots I found these really resonated with my understanding of feedback interaction now. I saw the similarity that I'm just trying to keep a specific sound.'' (\anon[Author1]{Yichen}, D4, L7--11)

``an alignment between our approaches, you've got the \ldots noise tap input. My equivalent of that is like more or less constant tapping or changes on the touchpad, which I can do quite subtly and carefully, which just effectively stops the RNN from interrupting, or it can only make short interruptions. It keeps its context in a certain place.'' (\anon[Author2]{Charles}, D4, L111)
\end{quote}
While our approaches and our instruments' affordances were different, we were both engaging in maintenance of the feedback state of our systems.

The reflection and diffractive reading process enabled us to engage with strategies we adopted as part of our feedback musicianship. 
While our strategies were quite different we identified their application as maintaining the feedback instruments' state: 
Supporting it, as in \anon[Author1]{Yichen}'s noise injections, or restricting it, as in \anon[Author2's]{Charles'} funnelling through tapping or touch gestures.
The goal of this maintenance was to support our instruments' agency within our performances, guiding them, through encouragement or restriction, to an artistically rewarding state.

\subsection{Navigating agency is a core skill for feedback musicianship}
\label{theme:nav-agency}

Our Phase 1 reflection prompts framed instrument behaviour as a binary between leading and responding. 
While this framing seemed appropriate in terms of accepted music interactions, such as exchanging solos in a jazz ensemble, our reflections and discussions led us to reject this as not aligning with the experience of our feedback instruments.
Rather, we recognised a more continuous space for agency, and that navigating the agency of feedback instruments was a skill that we were strengthening over the five performances.

It was clear in our discussions that we were uneasy with the framing of our instruments as either leading or responding and that our feedback instruments' agency may be better defined as a continuous spectrum.
\begin{quote}
``I started to question whether asking about leading moments was even a relevant question\ldots feeling like the instruments were leading rather than responding\ldots It was that the sense of control was more dispersed between the instruments and the musicians. So those were the moments where I thought that things were really working and the concept of the feedback instruments was paying off.'' (\anon[Author2]{Charles}, D2, L190--195)

``\anon[Author2]{Charles} noted that this unexpected moment is more of a spectrum and that the instrument rarely seem to respond rather they're either in a stable state or deviating from that. I totally agree with that\ldots having some unexpected moment in my practice seems to be the start of going to prepare for the feedback environment. I think a spectrum is a good word because you really don't know. It's all the times you're kind of managing the risk or preparing for these moments, but it might not show up.'' (\anon[Author1]{Yichen}, D3, L95--97)
\end{quote}
In D2 and D3 we became more confident in the spectrum framing, and started to discuss how we would manage or prepare for changes in the instruments' agency.

In the final dialogue, we identified this activity as \emph{navigation}.
\begin{quote}
``I felt better able to navigate the change of agency and control from distinct to controlled to diffused and dissolved. So I felt like I had some more ability to let the instrument take over or let the control become more diffuse at moments.'' (\anon[Author2]{Charles}, D5, L119--121)
\end{quote}
Beyond the techniques described in Section~\ref{theme:agency-maintenance}, this was a skill involving observation of our instruments' behaviour in the ensemble context.
The skill was developed over the multiple performances:
\begin{quote}
``Did you feel like you had better trust over your system by the time we got to this final performance?'' (\anon[Author2]{Charles}, D5, L67)

``Yes. I think I had a better understanding, as you said, I have lots of activities. I was trying to introduce a constraint or an adjustment of the input sound. I had a better understanding of ways I can work with this out of control moments.'' (\anon[Author1]{Yichen}, D5, L68--69)
\end{quote}
The skill of navigating agency seemed not just to apply to our own instrument but across the ensemble:
\begin{quote}
``we, in this recording session, started to feel more confident about each other's performance as well as our own. And so we were already feeling a bit more aligned in our expectations for each other's instruments.'' (\anon[Author2]{Charles}, D4, L91--92)
\end{quote} 

As \anon[Author1]{Yichen} and \anon[Author2]{Charles} noted above, our developing skill gave us further confidence in the feedback instruments' behaviour. 
In the earlier D3 we had established that confidence led to trust, and more exploratory performing:
\begin{quote}
``we were quite considered about where we were going to take over control of the instrument\ldots we had better feedback information from the instruments, or more practiced feedback information\ldots we also had a bit more trust in the instruments in terms of letting them operate independently to see what would happen.'' (\anon[Author2]{Charles}, D3, L14--19)
\end{quote}
The idea that this skill enabled better co-exploration with the feedback instruments is supported by \anon[Author1]{Yichen}'s reflection on P5:
\begin{quote}
``reflecting on out-of-control moments, I felt these would be leading moments, that before having it in control through my interaction to change the pitch, and LFO for the rhythm and movement changes over time, I allowed it to continue and observed what the AI would generate on its own, whether it could produce sounds beyond my anticipation.'' (\anon[RA1,]{RY,} L156)
\end{quote}

Our reflections and discussions complicated the Phase 1 framing of our instruments' behaviour as either leading or responding.
In our performances, the instruments did not appear to match either description strongly, rather their agency shifted through different moments of stability and deviation.
We responded to those shifts through a skill that we ended up calling navigation which became strengthened over the five performances.
Navigating agency was applied to our own instruments and across the ensemble. This skill yielded confidence to trust our instruments, allowing them to operate independently, giving us access to sounds that we couldn't anticipate.
This finding connects with the themes discussed in sections~\ref{theme:shared-exploration} and~\ref{theme:agency-maintenance}. Navigating agency enabled our shared exploration described earlier so we describe it as a core skill for feedback musicianship.

\section{Discussion}

The five themes generated from our diffractive dialogue articulate feedback musicianship in terms of enacted relationships, skills and techniques, and a process of discovery. 
In this section we connect the themes to wider understandings of these topics and reflect on our understanding of diffractive dialogue in practice-based research.

\subsection{The enacted relationships in feedback musicianship}

We argue that control and agency are not held by either performer or instrument, but are enacted relationships shifting across performances and musical moments (sections~\ref{theme:control-relational} and~\ref{theme:textures-enact}).
This supports findings from \citet{feedback:magnusson2022} that feedback systems exhibit a relational agency shaped by musicians' perceptions. We observed mechanisms similar to playing with, against, or being played by, as found by those researchers. 
Our project surfaced changes in feelings of control and agency over time; the relationship with our different instruments was enacted across the performances.
Strikingly, the understanding of control held by each performer reversed across the performances. \anon[Author2]{Charles} initially restricted Agentier from excessive divergence, while \anon[Author1]{Yichen} cajoled Sù into exploration for the instrument to reach a useful state, but these positions swapped by the final performances. 
This evidence supports treating control as enacted within the entangled practice, not held by performer or instrument.

In Paul Stapleton's practice, as articulated by \citet{kiefer2025feedbackmusicianship}, feedback was a ``means to distribute agency across a variety of actors within a musical ecosystem'', an enactivist approach.
\citet{playingfeedback:mudd2023} framed feedback musicianship as ``letting go of intention, embracing an entangled agency''.
Similarly we found in Section~\ref{theme:textures-enact} that different musical textures (as well as the instruments and performers) enacted co-agency on our feelings of control, leading to rewarding \emph{feedback moments}.
\anon[Author2]{Charles} reflected that the dense musical textures supplied a trusting environment, allowing him to distribute agency. As for sparse musical texture, \anon[Author2]{Charles} related it to tight control between two instruments, while \anon[Author1]{Yichen} associated it with a loss of control that needed to be regained  through active influence.
The single texture carrying two readings was a diffractive finding indicating that texture is the substrate through which control can be enacted by performers.
The enacted performer-instrument relationship in these musical textures formed a co-agency that stimulates unique musical controls for working feedback moments.

Our findings build on previous work that has identified the relational agency between performer and instrument. We found that agency was constructed relationally through the characteristics of the intelligent feedback instruments themselves, with other musicians, with the musical textures during performances, and was dynamic over our performance practice. This justifies our finding that navigating this agency is a core skill for feedback musicianship (Section~\ref{theme:nav-agency}).

\subsection{Skills and techniques of feedback musicianship}

We found in Section~\ref{theme:shared-exploration} that both intelligent feedback instruments are stateful and require a shared-exploration to enter a productive feedback environment.
Beyond acoustic feedback, our instruments contained AI models' realtime states within their trained behaviour, adding further layers of state to manage.
This relates to what \citet{Borgo:2005kk} called ``exploring and negotiating uncertainties together'' (p. 14), a defining part of music improvisation.
Exploring uncertainty was not just preferable, but necessary to access the productive state of our instruments.
We argue that this necessary, shared, performed exploration of dynamic state distinguishes feedback musicianship.

Skills such as van Eck's ``tuning'' and Stapleton's ``embodied way of knowing'' have been previously reported by \citet{kiefer2025feedbackmusicianship} as ways to engage with feedback.
Our work also surfaced techniques for maintaining the stateful behaviours of our instruments (Section~\ref{theme:agency-maintenance}).
Funnelling and injection were identified as techniques to restrict and activate feedback states.
\anon[Author2's]{Charles'} funnelling, guiding Agentier by constant adjustment, managed moments where it led in unwanted directions.
\anon[Author1]{Yichen}'s injection, re-energising the feedback loop with a noise tap, recovered Sù from unresponsive or out-of-control moments.
These techniques fall outside traditional musicianship: they treat our feedback instruments as co-agents, not controlled objects.
By later performances we had grown confident in these techniques. 
We trusted our instruments more, accessed new states and sounds, and chose better moments to intervene.
We noted that our instruments displayed a spectrum of agency, \textit{``they're either in a stable state or deviating from that''} (\anon[Author1]{Yichen}, D3), and we learned to navigate it.
Thus, the overarching skill of our feedback musicianship was \emph{navigating agency}. 
Observing our instruments evolve and engaging with their uncertain, dynamic states aligns with \citepos{eldridge-computer-musicking:2022} ``complexity literacy''; navigating agency goes further, enabling us to not just read but guide the changing agency of our instruments across performances.

\subsection{Design implications for intelligent feedback instruments}

This research has focussed on the experience of working with specific feedback instruments, rather than the process of building or iterating on new instruments by authors such as \citet{eldridge-feedback-cello:2017}, \citet{ulfarsson-halldorophone:2018} or \anon[Anonymised~\citep{anon}]{Martin~\citep{martin-opening-design-space:2026}}.
Even so, the themes identified in our results suggest design considerations for intelligent feedback instruments, contributing to the gap in digital feedback knowledge identified by \citet{playingfeedback:mudd2023}.

Our findings surfaced the feedback environment as a design target; a rewarding state that is necessarily difficult to enter. This opposes the NIME tradition for ``relatively easy''~\citep{Wessel:2002ab} instruments. 
Instead, we align with practices where the state of the instrument, environment, and performer are intrinsic, as in the Feedback Cello~\citep{eldridge-feedback-cello:2017} and Halldorophone~\citep{ulfarsson-halldorophone:2018}.
Intelligent feedback instruments should be designed for statefulness, under the expectation that short-cuts to the feedback environment are not rewarding. 
\citepos{sophtar:visi2024} Sophtar shows how feedback controls can be exposed within an embedded AI instrument; despite controls, we found that entering and maintaining the feedback environment had to be learned. Affordances for these techniques could be designed into feedback DMIs.

Feedback musicianship, in our practice, involved a spectrum of co-agency, and a necessity for building skills.
The relational control of feedback instruments pushes against the framing of DMIs as controllers~\citep{Wessel:2002ab}. 
Traditional terms such as ensembles, or leader and responder, do not articulate the changing relational agency that we observed. 
Successful intelligent feedback instrument designs should address the spectrum of co-agency that emerges in feedback musicianship and acknowledge the skills required. 
Navigating agency with one's instrument is not a typical aspect of musicianship but a similar skill to \citepos{eldridge-computer-musicking:2022} ``complexity literacy''.

\subsection{Generating concepts with diffractive dialogue}

This research borrowed the diffractive dialogue methodology introduced by \citet{morrison2024entangle}. In our project, the dialogue was spoken in real-time, rather than written.
In our process, the dialogue was the main site for creating meaning in this research.
Our contributions to a vocabulary for feedback musicianship came from dialogue, not a later coding process.
Examining the performances in sequential diffractive dialogue surfaced development and even reversal of views, such as the flipped interpretation of control as stability or diversity in Section~\ref{theme:control-relational}.
The process allowed us to react against the initial prompt framings such as the binary leading/following frame in Section~\ref{theme:nav-agency}.
In the context of practice-based first-person research, our diffractive dialogue served as a natural way to make sense of our feedback musicianship.
The process of discussing the changing moments of contrast or alignment created an understanding of our practice that was not evident from the individual reflections.

\subsection{Limitations}

The findings in this study were sourced specifically from the duet musical practice of the first two authors. While our practice with intelligent musical instruments has extended for more than two years, the performances studied in our diffractive dialogue span a period of two months. Further insights may be available by expanding the length of time considered or studying larger groups of musicians.
While our work concerned the feedback interactions in intelligent musical instruments we did not consider the design of these instruments in depth, or how the AI models might be trained. \citepos{stefannsdottir2025altered} work is a good example of how AI model training can be integrated into an artistic research context for feedback musical instruments.

\section{Conclusion}

This research uses a duet performance practice using intelligent feedback instruments, framed by agential realism, to develop our understanding of the relationship between performers and instruments in feedback musicianship.
Previous understandings of feedback musicianship primarily involve case studies of individual experience.
Uniquely, in this work, we analyse two musicians in one practice with different feedback instruments; one operating on enacted audio data, the other on embodied gestural data.
We apply a diffractive dialogue approach to read meaning through each other's reflections on five performance experiences.

Our diffractive dialogue found that control and agency in feedback musicianship are relational rather than fixed, shifting with the instrument's behaviour, the co-performer, and the performers' evolving intentions. 
Our positions on the meaning of control reversed across the five performances even though the instruments themselves did not change. 
Reaching a productive state with these instruments was itself an exploratory act, a process of working together into what we called the \emph{feedback environment}. 
Once there, we maintained our instruments' agency through specific techniques: \anon[Author1]{Yichen} \emph{injected} noise into Sù to wake the model, while \anon[Author2]{Charles} \emph{funnelled} Agentier with constant input to keep it from escaping. 
Across the performances we developed the skill of \emph{navigating agency}, which displaced the leading-versus-responding binary in favour of a continuous spectrum of control states. 
These five concepts, \emph{feedback moments}, \emph{feedback environment}, \emph{funnelling}, \emph{injection}, and \emph{navigating agency}, contribute to the emerging vocabulary of feedback musicianship.

Our themes argue that performing with intelligent feedback instruments calls for a different musicianship than the tool-oriented or leader-and-responder framings inherited from earlier DMI and ensemble research. 
Feedback musicianship demands navigation of shifting relational agency where musical textures enact feedback moments with performers and instruments.
We offer implications for intelligent feedback instrument design, prioritising statefulness and skill building, alongside a refined procedure for spoken diffractive dialogue in which the dialogue was the site where meaning was generated. 
We suggest that the relational agency surfaced here is not a property that can be designed in, but one that must be worked for through practice. 
Applying agential realism to feedback musicianship in particular, and human-AI creative collaboration in general, may help surface this relational agency in other practices where it is similarly worked for and rewarding.

\paragraph{AI Acknowledgement}

We acknowledge the thoughtful use of Claude AI via Claude Code in assisting with data analysis and editing of this manuscript. Consistent with our topic of creative human-AI collaboration, we found navigating the entangled agency of AI in research can lead to rewarding feedback moments.

\begin{acks}
Yichen Wang's instrument 溯 (Sù) was developed during a research internship program at Dolby ATG Group, Sydney, Australia in 2025.
Aspects of the instrument 溯 are subject to a patent application with Dolby ATG.
We thank Benjamin Southwell and Jeroen Breebaart for additional guidance in the development of 溯 (Sù) at Dolby.
This research was partially supported by the Research Council of Norway through the MishMash Centre for AI and Creativity, Project Number 357438.
\end{acks}

\bibliographystyle{apacite}
\bibliography{references}

\end{document}